\documentclass[a4paper,fleqn,usenatbib]{mnras}

\usepackage[T1]{fontenc}
\usepackage{aecompl}
\usepackage{graphicx}	
\usepackage{amsmath}	
\usepackage{amssymb}	
\usepackage{bm}
\usepackage{makecell} 
\usepackage{array}

\title[First H\,{\normalsize I} Detection in a Strong Spiral Lens]{The First Detection of Neutral Hydrogen in Emission in a Strong Spiral Lens}

\author[A. Lipnicky et al.]{
Andrew Lipnicky,$^{1,3}$\thanks{E-mail: awl6964@rit.edu (AL)}
Sukanya Chakrabarti,$^{1}$
Melvyn C. H. Wright,$^{2}$ 
Leo Blitz,$^{2}$
\newauthor Carl Heiles,$^2$
William Cotton,$^{3}$
David Frayer,$^{4}$
Roger Blandford,$^{5}$
Yiping Shu,$^{6}$
\newauthor and Adam S. Bolton$^{7, 8}$
\\
$^{1}$School of Physics and Astronomy, Rochester Institute of Technology, Rochester, NY 14623, USA\\
$^{2}$Department of Astronomy, University of California, Berkeley, CA 94720, USA\\
$^{3}$National Radio Astronomy Observatory, 520 Edgemond Rd, Charlottesville, VA 22903, USA\\
$^{4}$National Radio Astronomy Observatory, P.O. Box 2, Green Bank, WV 24944, USA\\
$^{5}$KIPAC, Stanford University, P.O. Box 20450, MS 29, Stanford, CA 94309, USA\\
$^{6}$National Astronomical Observatories, Chinese Academy of Sciences, 20 A Datun Rd, Chaoyang District, Beijing 100012, China\\
$^{7}$National Optical Astronomy Observatory, 950 N Cherry Ave, Tucson, AZ 85719, USA\\
$^{8}$Department of Physics and Astronomy, University of Utah, 115 South 1400 East, Salt Lake City, UT 84112, USA\\
}

\date{Accepted 2018 February 01. Received 2017 December 04; in original form 2017 July 22.}

\pubyear{2018}

\begin{document}
\label{firstpage}
\pagerange{\pageref{firstpage}--\pageref{lastpage}}
\maketitle

\begin{abstract}
We report H\,\textsc{i} observations of eight spiral galaxies that are strongly lensing background sources. Our targets were selected from the Sloan WFC (Wide Field Camera) Edge-on Late-type Lens Survey (SWELLS) using the Arecibo, Karl G. Jansky Very Large Array, and Green Bank telescopes. We securely detect J1703+2451 at $z = 0.063$ with a signal-to-noise of 6.7 and $W_{50}=79\pm13$ km s$^{-1}$, obtaining the first detection of H\,\textsc{i} emission in a strong spiral lens.  We measure a mass of $M_\text{H\,\textsc{i}}= (1.77\pm0.06^{+0.35}_{-0.75})\times10^9 \text{ M}_{\odot}$ for this source. We find that this lens is a normal spiral, with observable properties that are fairly typical of spiral galaxies.  For three other sources we did not secure a detection; however, we are able to place strong constraints on the H\,\textsc{i} masses of those galaxies. The observations for four of our sources were rendered unusable due to strong radio frequency interference.

\end{abstract}



\section{Introduction}
Gravitational lensing has been used extensively as a tool for constraining the amount of dark matter contained within galaxy clusters.  With the advent of large scale surveys, it has now been found that many individual galaxies are also strong gravitational lenses and can be used to provide constraints on their dark matter haloes.  At these small scales, the agreement between cosmological simulations and observations is not yet as secure as on large scales \citep{Klypin:1999aa, Dutton:2007aa, Weinberg:2015aa}.  For example, with the SWELLS sample \citep[Sloan WFC (Wide Field Camera) Edge-on Late-type Lens Survey,][]{Treu:2011aa, Brewer:2012aa}, using a set of strong-lens spiral galaxies selected in the same manner as the larger SLACS lens survey \citep[Sloan Lens ACS (Advanced Camera for Surveys);][]{Bolton:2008aa} from within the spectroscopic data set of the Sloan Digital Sky Survey \citep[SDSS;][]{York:2000aa}, studies have attempted to break the ``Disk-Halo Degeneracy,'' a problem that arises from the fact that rotation curves can be fit equally well by having either a high mass halo and low mass disk or vice-versa.  The problem can be alleviated by identifying strong spiral lenses and combining gravitational lensing analysis with stellar kinematics \citep{Dutton:2011aa, Treu:2011aa, Brewer:2012aa}. 

Other work suggests that the Missing Satellites problem, the discrepancy between the number of observed and simulated subhaloes, may only apply to the Local Group.  \citet{Vegetti:2012aa} analyzed surface brightness anomalies in gravitationally lensed images of an elliptical deflector at $z=0.8$ and argued that their derived constraints on cold dark matter (CDM) substructure are consistent with simulations.  However, gravitational lensing analyses have concentrated mainly on deflectors that are ellipticals and there are indications that the distribution of substructure may be intrinsically different in ellipticals than in spirals of the same mass \citep{Nierenberg:2012aa}.  These discrepancies, along with currently overly simplistic models for lensed galaxies (Treu et al. 2011), call for an independent measurement of dark matter distribution.   

Alternate probes of the dark matter distribution, whether of the potential or dark matter substructure, can be obtained by analyzing stellar tidal debris \citep{Johnston:1999aa}, velocity asymmetries in the stellar disk \citep{Widrow:2012aa, Xu:2015aa}, or disturbances in outer H\,\textsc{i} disks \citep{Chakrabarti:2009aa, Chakrabarti:2011aa, Chakrabarti:2011ab, Chang:2011aa, Chakrabarti:2013aa}. The gas disk is a particularly sensitive probe of the dark matter distribution as it is kinematically colder than the stars, and extends out to several times the optical radius \citep{Wong:2002aa}.  Recent extragalactic H\,\textsc{i} surveys \citep{Giovanelli:2005aa, Catinella:2010aa, Fernandez:2016aa} and future prospects with the Square Kilometer Array are especially promising \citep[SKA;][]{Giovanelli:2016aa} as they open a new window for H\,\textsc{i} studies beyond the Local Volume.

The work done by \citet{Chakrabarti:2009aa, Chakrabarti:2011aa, Chakrabarti:2011ab, Chang:2011aa} has shown that one can constrain the mass, current radial distance, and azimuth of a galactic satellite by finding the best-fit to the projected gas surface density of an observed galaxy.  Then by performing and searching a set of hydrodynamical simulations, a best-fit to the observed data can be obtained.  This method is called "Tidal Analysis". \citep{Chakrabarti:2009aa, Chakrabarti:2011aa}. \citet{Chakrabarti:2011ab} applied this method to M51 and NGC 1512, both of which have optical companions about $1/3$ and $1/100$ the mass of their hosts respectively.  Using the Tidal Analysis method, they found that the masses and relative positions of the satellites in both systems were accurately recovered.  Moreover, the fits to the data were found to be insensitive to reasonable variations in choice of initial conditions of the primary galaxy or orbital inclination and velocity of the satellite.  The advantage with this method is that it can be used to find dark satellites, as long as they are at least 0.1\% the mass of the primary. 

H\,\textsc{i} is an ideal tracer of the outer parts of a disk galaxy.  A successful detection of H\,\textsc{i} in a spiral lens will
ultimately enable us to make H\,\textsc{i} images that we can analyze in a similar manner to our earlier work \citep[e.g.][]{Chakrabarti:2011ab} to characterize dark matter substructure.  Furthermore, with a H\,\textsc{i} detection alone, one can obtain an estimate of the velocity dispersion of the gas that can be compared to the velocity dispersion from lensing.  We expect that with forthcoming H\,\textsc{i} maps of these spiral lenses, Tidal Analysis can be used to independently constrain the dark matter distribution of spiral galaxies and may provide accurate priors for more precise strong lensing models.  Lensing and Tidal Analysis probe different regions of galaxies as gravitational lensing is sensitive to the inner regions and Tidal Analysis is sensitive to the outer regions where the gas disk is most easily disturbed by passing satellites. Therefore, the combination of methods can provide a more complete picture of substructure in a galaxy, and can potentially help to resolve some of the outstanding problems with the CDM paradigm on galactic scales.  The SWELLS sample is a sufficiently low redshift sample of strong spiral lenses ($z_\text{avg}\sim0.1$) that enables these two methods to be combined for the first time.  The galaxies in this sample also show visible signs of interaction with satellite galaxies and therefore probable substructure. 

In the future, we plan to obtain H\,\textsc{i} maps of this sample for which constraints on past interactions and substructure can be obtained using Tidal Analysis. At the same time, analysis of surface brightness anomalies in the lensed image can be used to also constrain substructure \citep[e.g.][]{Vegetti:2012aa}.  Thus, we should be able to compare and contrast two independent methods of characterizing substructure. As a first step, here we investigate if the target spiral lens galaxies have sufficient H\,\textsc{i} masses so that follow-up, high resolution H\,\textsc{i} mapping will be successful. 

The paper is organized as follows: In Section 2 we discuss our sample selection. In Section 3 we discuss the observations and data reduction. In Section 4 we show our results and determine H\,\textsc{i} masses for our targets. In Section 5 we discuss the results and their implications. Finally, in Section 6 we conclude.  All distance-dependent quantities in this work are computed assuming $\Omega = 0.3$, $\Lambda = 0.7$, and $H_0 = 70$ km s$^{-1}$.

\section{Sample selection} 
The galaxies we observed lie at low redshift (the average redshift of the SWELLS sample is $z_\text{avg} \sim 0.1$) but still at the very edge of what is currently possible for H\,\textsc{i} mapping observations.  Recently, \citet{Fernandez:2016aa} presented an H\,\textsc{i} map of a massive spiral ($M_\text{H\,\textsc{i}}=2.9\times10^{10} \text{M}_{\odot}$) at $z=0.376$ observed with the Karl G. Jansky Very Large Array (VLA) for 178 hours and obtained a signal-to-noise of $S/N=7$ for their integrated spectrum, although the structure in the outskirts is uncertain. This is the highest redshift H\,\textsc{i} emission map obtained to date. Previously, that record was held by \citet{Donley:2006aa}, who observed HIZOAJ0836-43 with the ATCA for $\sim2\times12$ hours which is a large disk galaxy at $z=0.036$ with $M_\text{H\,\textsc{i}}\sim7\times10^{10}\text{ M}_{\odot}$ and a H\,\textsc{i} diameter of $\sim$130 kpc. 

\begin{figure}
\centering
\includegraphics[width=\columnwidth]{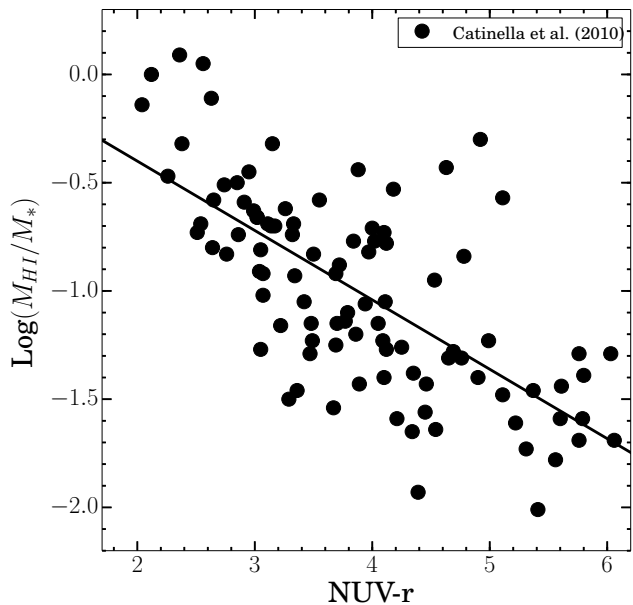}
\caption{A clear correlation exists between gas fraction, $\log(M_\text{H\,\textsc{i}}/M_{\star})$, and the NUV-\textit{r} color of a galaxy. From a linear fit to this data (black line), we get a broad indication of the expected H\,\textsc{i} mass for our galaxies. Data from Data Release 1 of \protect\citet{Catinella:2010aa}.}
\label{NUV_r_DR1}
\end{figure}

\citet{Treu:2011aa} have derived stellar masses for the SWELLS sources based on multiband photometry and stellar population synthesis models. Two common initial mass functions (IMFs) were used, the \citet{Chabrier:2003aa} and \citet{Salpeter:1955aa} IMFs. \citet{Brewer:2012aa} carried out a gravitational lensing analysis of the SWELLS sources. By fitting a singular isothermal ellipsoid (SIE) mass model for the deflector, they were able to use the Einstein angle and axis ratio to measure a velocity dispersion. Thus, they have provided velocity dispersions and stellar masses for the SWELLS sample; however, the H\,\textsc{i} masses were not measured. Unfortunately, galactic gas fraction, $M_\text{H\,\textsc{i}}/M_{\star}$, is not tightly correlated with stellar mass \citep{Catinella:2010aa}; therefore, we must pre-observe objects of interest with large single dish telescopes in order to ascertain their total mass in H\,\textsc{i}.  In order to narrow our search, we estimate the expected H\,\textsc{i} mass of each galaxy. From the \textit{GALEX}-Arecibo-SDSS survey (GASS) of $\sim$1000 galaxies studied between redshifts $0.025 < z < 0.05$ with $M_{\star} > 10^{10} \text{ M}_{\odot}$, \citet{Catinella:2010aa} found that $M_\text{H\,\textsc{i}}/M_{\star}$ was most clearly correlated with the NUV --- \textit{r} color, although even here there is a large dispersion (Fig. \ref{NUV_r_DR1}). We estimate the H\,\textsc{i} mass of each source by using this linear relationship between gas fraction and NUV --- \textit{r} color. Furthermore, we estimate the expected flux density of each source by assuming a velocity width and profile shape for the H\,\textsc{i} 21-cm spectral line. The average stellar mass of the SWELLS sources is $\log(M_\star)\sim10.7$. Based on the velocity widths of similar galaxies from the GASS survey, we assume a velocity width of 300 km s$^{-1}$. Also, since most of our sources are inclined to some degree, a double-horned line profile is expected. For simplicity, we assume a simple top-hat velocity profile shape. Because of the large scatter in the relation shown in Fig. \ref{NUV_r_DR1}, our predictions are only rough estimates of the actual H\,\textsc{i} mass.

\begin{table*}
\caption{Data for the SWELLS targets which were studied in this investigation. Column 1 gives the source ID for each object which corresponds to the deflector or nearby object that we are interested in H\,\textsc{i} mapping. Columns 2 and 3 give the J2000 RA and Dec of each source, respectively. Column 4 gives the approximate SDSS redshift of the deflector object which we are interested in observing. Column 5 gives the stellar velocity dispersion as measured from fits to the stellar continuum \protect\citep{Bruzual:2003aa}. Column 6 gives the velocity dispersion as measured from the gravitational lensing analysis presented in \protect\citet{Brewer:2012aa}. Column 7 gives the stellar masses of each galaxy, compiled from \protect\citet{Treu:2011aa} and \protect\citet{Brewer:2012aa} and are the averaged disk masses given Chabrier and Salpeter IMFs. Column 8 gives the NUV --- \textit{r} color of each source based on data from \textit{GALEX} and SDSS observations. Column 9 gives the gas fraction of each source, $\log(M_\text{H\,\textsc{i}}/M_\star)$, and is calculated from a fit to the gas fraction versus NUV --- \textit{r} color from \protect\citet{Catinella:2010aa} as displayed in Fig. \ref{NUV_r_DR1}.  Column 10 gives the predicted H\,\textsc{i} mass of each source based on the fit to the \protect\citet{Catinella:2010aa} data. Finally, Column 11 gives the expected signal in mJy of each source, calculated by rearranging equation \ref{approxMHI} and assuming a velocity width of 300 km s$^{-1}$ and a top-hat line profile shape.}
\begin{center}
{\small
{\renewcommand{\arraystretch}{1.3}
\begin{tabular}{ccccccccccc}\hline\hline
\makecell{Source ID\\\\(1)} &\makecell{RA\\(deg)\\(2)}&\makecell{Dec\\(deg)\\(3)} & \makecell{$z_d$\\\\(4)} & \makecell{$\sigma_\text{SDSS}$\\(km s$^{-1})$\\(5)}	& \makecell{$\sigma_\text{SIE}$\\(km s$^{-1}$)\\(6)} & \makecell{$\log(M_\star)$\\(M$_{\odot}$)\\(7)} & \makecell{NUV-\textit{r}\\\\(8)} & \makecell{$\log(M_{\text{H}\,\textsc{i}}/M_\star)$\\\\(9)} & \makecell{$\log(M_\text{H\,\textsc{i}})_\text{exp}$\\(M$_{\odot})$\\(10)} & \makecell{$S_\text{exp}$\\(mJy)\\(11)} \\ \hline
J0329-0055 & 52.48734	&-0.93151&0.1062 & 94$\pm$25		& \ldots			&10.30   & 2.90  & -0.7        & 9.6      & 0.28 \\
J0841+3824 &130.37004	&38.40381&0.1160 & 217$\pm$18		& 251.2$\pm$4.4	&11.35   & 3.59  & -0.9        & 10.4     & 1.57 \\
J1037+3517 &159.43764	&35.29194&0.1220 & 243$\pm$26	& \ldots	&10.44   & 4.01  & -1.0        & 9.4     & 0.13 \\
J1103+5322 &165.78421	&53.37450&0.1580 & 235$\pm$29 	& 222.8$\pm$3.1	&10.84   & 4.51  & -1.2        & 9.6      & 0.13   \\
J1111+2234 &167.86674	&22.58072&0.2223 & 236$\pm$24		& 227.9$\pm$1.5	&10.99   & 4.84  & -1.3        & 9.7      & 0.07    \\
J1117+4704 &169.39742	&47.06873&0.1690 & 197$\pm$22		& 217.5$\pm$2.5	&11.05   & 4.37  & -1.2        & 9.9      & 0.21    \\
J1135+3720 &173.77867	&37.33997&0.1620 & 211$\pm$26	& 206.3$\pm$10.3	&11.10   & 3.62  & -0.9        & 10.2     & 0.44  \\
J1703+2451 &173.77867	&37.33997&0.0630 & 198$\pm$33		& 189.7$\pm$2.2	&10.81   & 5.81  & -1.6        & 9.2      & 0.30  \\ \hline
\end{tabular}}
}
\end{center}
\label{SWELLS-sources}
\end{table*}

In an effort to combine gravitational lensing with Tidal Analysis, we chose a set of galaxies that were most likely to have visible substructure.  We chose ``Grade A'' (unambiguous multiple images reproduced by a relatively simple lens model) strong spiral lenses \citep{Treu:2011aa, Brewer:2012aa} as well as two ``Grade B'' (probable lensing structure) lenses that have plausible lens models (Bolton; private communication) from the SWELLS sample.  All galaxies in our sample are star-forming spirals with strong $H_\alpha$ emission with various lensed features.  They have varying inclinations, redshifts, disk-to-bulge ratios, masses, and clarity of lensing features. These sources have been observed in NUV by the \textit{Galaxy Evolution Explorer} \citep[\textit{GALEX};][]{Martin:2005aa} and in $r$ by SDSS \citep{Alam:2015aa}. Other factors which contributed to our sample selection included the ability to observe them with the Arecibo and/or Green Bank telescopes, the expected H\,\textsc{i} mass, expected signal, the probability of substructure, and the Radio Frequency Interference (RFI) environment at the redshifted frequency. Due to redshift, the frequency location of the 21 cm emission is a particularly hostile portion of the radio spectrum. Beyond redshift of $z\sim0.05$ ($\nu\lesssim$1350 MHz) the spectrum is heavily populated by satellite communications, airplane navigation, airport, military, and defense radar, cellular communications, and other unknown sources of noise. Shown in Table \ref{SWELLS-sources} is information on the lenses from the SWELLS survey that were observed in this investigation, their predicted H\,\textsc{i} masses, and the expected signal. Our predictions for H\,\textsc{i} masses are based off the linear fit seen in Figure \ref{NUV_r_DR1}. Since this relation is wide and uncertain, we do not place any errors on estimated values in Table \ref{SWELLS-sources} to highlight their uncertain nature.

\section{Observations and Data Reduction}

J1037+3517 and J1111+2234 were observed using the Arecibo Telescope with the L-wide receiver on 2014 April 23 and 2014 May 14 for a total of 5.25 hours. Targets were observed with a narrow bandwidth of 3.125 MHz to avoid RFI. This was separated into 2048 channels, two linear polarizations, and had a velocity resolution of $\sim$394 m s$^{-1}$. Observations were then carried out using the Green Bank Telescope (GBT) over the Fall semesters (August through January) of 2014 and 2016 for a total of 127.75 hours. The observations used the L-band receiver with the Versatile GBT Astronomical Spectrometer (VEGAS) backend in mode 15 which has a bandwidth of 11.72 MHz separated into 32,768 channels, and two linear polarizations. This yielded a velocity resolution of $\sim$87 m s$^{-1}$.  We used short, four minute scans with one second integration times to minimize the effect of RFI. For both the Arecibo and GBT observations, we used the observing method of position switching. 

All data reduction of the GBT observations was performed using \textsc{gbtidl} v2.10 \citep{Marganian:2013aa}. Due to the presence of significant RFI in all spectra, detailed data editing was necessary. This was done both automatically and by hand. Each one second integration was inspected automatically for the presence of wideband RFI (e.g. radar) and flagged if a signal stronger than five times the theoretical rms was present. The theoretical rms was computed using the radiometer equation

\begin{equation}
T_\text{rms} = \frac{\sqrt{2}T_\text{sys}}{\sqrt{N_\text{pol}\Delta f_\text{ch}t_\text{s}}},
\label{rms}
\end{equation}

\noindent where $T_\text{sys}$ is the system temperature, $N_\text{pol}$ is the number of polarizations, $\Delta f_\text{ch}$ is the channel bandwidth in hertz, and $t_\text{s}$ is the integration time in seconds. The extra factor of $\sqrt{2}$ comes from the combination of both the ON and OFF spectra, assuming that both spectra have the same noise.

Once RFI mitigation was performed, each observing session was combined into a single spectrum per polarization. Baseline fitting was limited to order $n\leq3$. It was found that third order fits produced the lowest rms; therefore, a third order polynomial baseline was subtracted from each spectrum. Then each source was combined into two master polarized spectra. This allowed us to check the spectra for RFI that mimicked H\,\textsc{i} signals as RFI is typically highly polarized. Finally, the two polarizations were combined into a final spectrum. This spectrum was then inspected by hand for narrowband RFI which was flagged. Narrowband RFI typically only affected a few channels, it was therefore blanked and interpolated over. The flux density scale was determined via online system temperature measurements with a blinking noise diode.  

For those spectra that were affected by noisy baselines, the measured rms (rms$_\text{obs}$) across the spectrum was compared to the theoretical rms (rms$_\text{theo}$). rms$_\text{obs}$ was measured using the \textsc{gbtidl} task \textsc{stats}. Integrations that showed rms$_\text{obs}>$ 3rms$_\text{theo}$ were discarded. 

Observations of J0329-0055 were performed during August 2013 with the VLA in ``C'' configuration for a total of 15.2 hours. The target was observed with a bandwidth of 64 MHz separated into 2048 channels at a velocity resolution of 8.2 km s$^{-1}$. Calibration and RFI excision of the VLA observations was performed using the standard \textsc{Obit} calibration pipeline \citep{Cotton:2008aa}. The standard calibrator, 3C138, was used for flux density, group delay, and bandpass calibrations and J0323+0534 was used for phase calibrations. The data were Hanning smoothed to $\sim$16.5 km s$^{-1}$ resolution before RFI excision. Then flagging was performed in both time and frequency domains at the 5$\sigma$ level by using a running mean. Calibration of the data was then repeated with the flagged data to further reduce the impact of RFI.

\section{Results}

\begin{figure*}
\begin{center}
\includegraphics[width=0.8\textwidth]{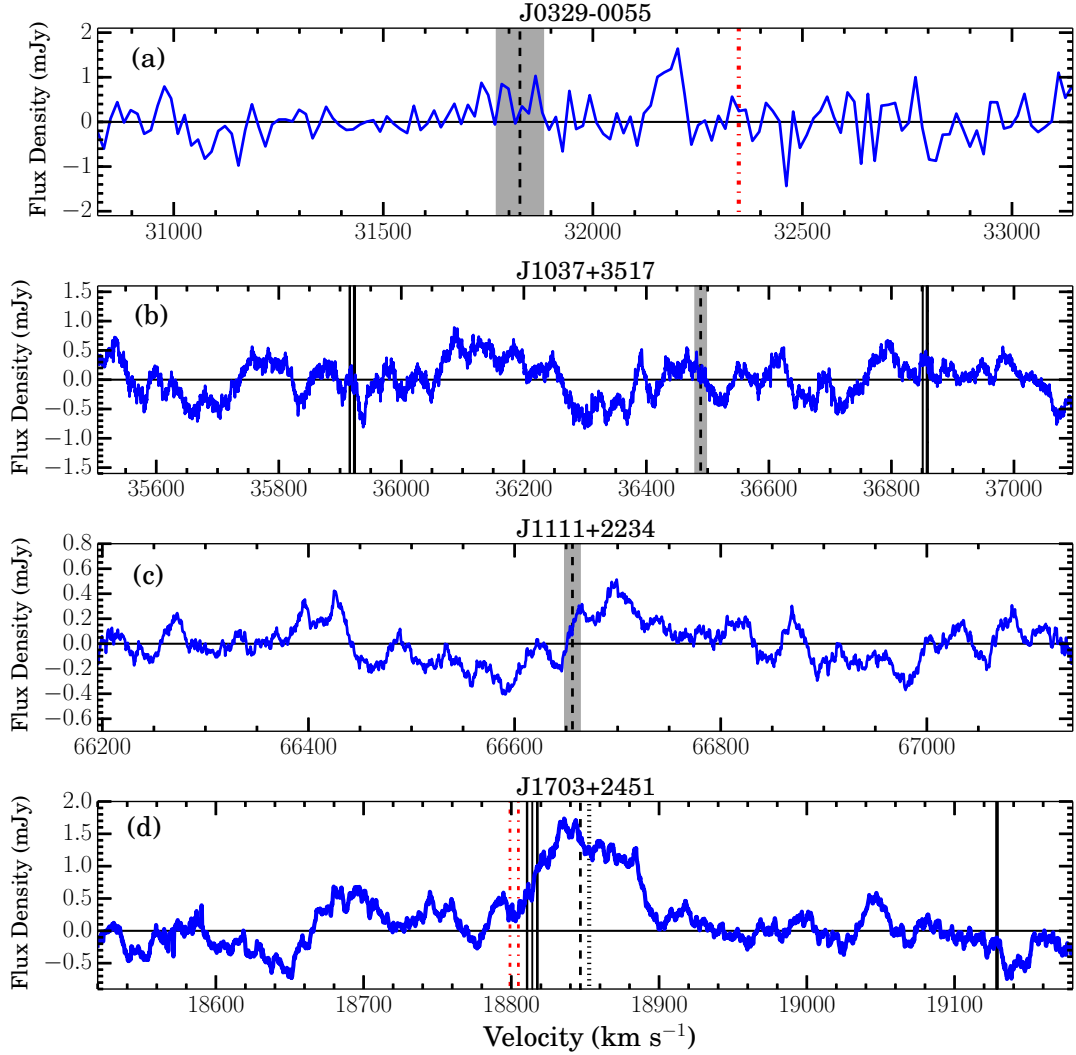}
\caption{In all panels solid black lines indicate parts of the spectrum that were blanked due to narrowband RFI, red dot--dashed lines indicate interloper galaxies that lie in the field of view of observations, dashed lines indicate the SDSS spectroscopically measured location of each source with the grey shaded regions marking their 3$\sigma$ error. Panel (a): this source was observed with the VLA and has been smoothed to a velocity resolution of $\sim$ 16.5 km s$^{-1}$. Panel (b) and Panel (d): these sources were observed with the GBT and both spectra have been Hanning smoothed and resampled, then boxcar smoothed to a velocity resolution of 20 km s$^{-1}$ and fit with a third order polynomial. Panel (c): this source was observed with Arecibo, it has been boxcar smoothed to a velocity resolution of 20 km s$^{-1}$ and fit with a third order polynomial.}
\label{spectra}
\end{center}
\end{figure*}

Of the eight objects observed, we have obtained one detection, three non-detection upper limits, and the data for four objects were completely lost due to RFI. We also report a serendipitous upper limit non-detection from the VLA. The detection we obtained of source J1703+2451 is shown in Fig. \ref{spectra}(d), the three non-detection upper limit spectra are shown in Fig. \ref{spectra}(a) -- (c), all our current mass estimates are displayed in Table \ref{SWELLS-sources}, and our observational results are displayed in Table \ref{rms_results}.

The H\,\textsc{i} mass of a galaxy can be obtained using the following relation:

\begin{equation}
M_{\text{H}\,\textsc{i}} \text{(M}_\odot)= \frac{2.356\times10^5}{1+z}D_\text{L}^2\int S  \text{d}v,
\label{MHI}
\end{equation}

\noindent where $z$ is the redshift to the source, $D_\text{L}$ is the luminosity distance to the galaxy in Mpc, and $\int S \text{d}v$ is the integrated flux of the H\,\textsc{i} signal in units of Jy km s$^{-1}$ \citep{Wild:1952aa, Roberts:1962aa}. 

For spectra that do not display a clear signal, we use an approximate form of equation \ref{MHI}:

\begin{equation}
M_{\text{H}\,\textsc{i}} {\text(M}_\odot) \approx \frac{2.4\times10^5}{1+z} D_\text{L}^2 S_\text{peak} W,
\label{approxMHI}
\end{equation}

\noindent where $S_\text{peak}$ is the peak of the signal in Jy and $W$ is the width of the signal in km s$^{-1}$. In the case of an upper mass limit, $S_\text{peak}$ is represented by three times the measured rms of the spectrum when smoothed to 20 km s$^{-1}$ and $W$ is assumed to be 300 km s$^{-1}$. This assumption is stated explicitly in all our mass estimates with an added $(\Delta v/300$ km s$^{-1})^{1/2}$ term.

\begin{table*}
\caption{Column 1 gives the Source ID. Column 3 displays the total usable time on source. Column 4 shows the total time spent observing the object including reference observations. Column 5 shows the percentage of the time on source that was lost due to RFI. Column 6 shows the theoretical rms in mJy computed using equation \ref{rms} in order to compare to Column 6 which displays the measured rms in mJy of the observed spectrum. Column 7 gives the mass that was determined from observations. Column 9 shows the telescope on which observations for that object were performed.}
\begin{center}
\begin{tabular}{cccccccc}\hline\hline
\makecell{Source ID\\\\(1)} & \makecell{$t_s$\\(hrs)\\(4)} & \makecell{$t_\text{tot}$\\(hrs)\\(5)}  & \makecell{RFI lost\\(\%)\\(6)} & \makecell{$\text{rms}_\text{theo}$\\(mJy)\\(7)}	&\makecell{$\text{rms}_\text{obs}$\\(mJy)\\(8)} & \makecell{$\log(M_\text{H\,\textsc{i}})_\text{obs}$\\(M$_\odot$)\\(9)} & \makecell{Telescope\\\\(10)}  \\ \hline
J0329-0055	& \textendash	& 15.2	& $>$50	& 0.31	& 0.38	& $< 10.37$	& VLA	\\ 
J1037+3517 	& 10.139	 	& 39.733 	& 49.0 	&0.20	& 0.32	& $<10.29$	 	&Arecibo/GBT\\  
J1111+2234	& 0.733		& 2.067	& 64.5	& 0.28	& 0.29	& 	$<10.78$	&Arecibo\\		
J1703+2451	&4.948		&10.8	& 8.4		&0.31	&0.27	& $9.25\pm0.01^{+0.08}_{-0.24}$	&VLA/GBT	\\ \hline 
\end{tabular}
\end{center}
\label{rms_results}
\end{table*}

\subsection{Notes on each source}

\subsubsection{J0329-0055:} Observations of this object were performed for a total of 15.2 hours. Due to RFI interference, more than 50 per cent of the data were flagged. After smoothing to a velocity resolution of $\sim$16.5 km s$^{-1}$, no signal was present at the expected source location (Fig \ref{spectra}(a)). A wide feature is seen centered at 29053.8 km s$^{-1}$ (1282.75 MHz); however, this does not correspond to any known source within the field of view. This frequency location also contains a known source of RFI corresponding to tethered aerostat radar system (TARS) pulsed radar from Ft. Huachuca, Arizona\footnote{https://science.nrao.edu/facilities/vla/observing/RFI/L-Band}. The measured rms of the spectrum around the expected signal location is $S_\text{rms} = 0.38$ mJy which yields an upper mass limit of $M_\text{H\,\textsc{i}} < 2.36\times10^{10}\times(\Delta v/300 \text{ km s}^{-1})^{1/2} \text{ M}_\odot$.

\subsubsection{J1037+3517:} This object was originally observed with Arecibo for 72 minutes where a tentative $2\sigma$ signal was present at the expected source location. However, due to the presence of strong RFI nearby, the source of the signal at the source location was not clear.  We then observed this object for a total of 39.7 hours with GBT; however, 49 per cent of the data was lost due to RFI. The final spectrum is shown in Fig. \ref{spectra}(b). The background RFI environment at this frequency location is higher than expected. From the smoothed spectrum, we observe no discernible signal at the expected source location and therefore estimate an upper mass limit using $S_\text{rms} = 0.32$ mJy which yields $M_\text{H\,\textsc{i}} < 1.96\times10^{10}\times(\Delta v/300 \text{ km s}^{-1})^{1/2} \text{ M}_{\odot}$.

\subsubsection{J1111+2234:} Observations of this source were performed for a total of 2.067 hours over two days; however, RFI strongly affected much of the observing time leading to a total usable observation time of 0.733 hours. The final spectrum is shown in Fig. \ref{spectra}(c). Since the Arecibo dish has three times the diameter of the GBT dish and thus nine times the collecting area, this allows for a low rms to be achieved in about a factor of ten less time on source. The final spectrum has a measured rms of $S_\text{rms}=0.29$ mJy, yielding an upper mass limit of $M_\text{H\,\textsc{i}} < 6.13\times10^{10}\times(\Delta v/300 \text{ km s}^{-1})^{1/2}\text{ M}_{\odot}$.

\subsubsection{J1703+2451:} 
\begin{figure} 
\includegraphics[trim={0cm 0.8cm 0cm 0cm},clip,width=\columnwidth]{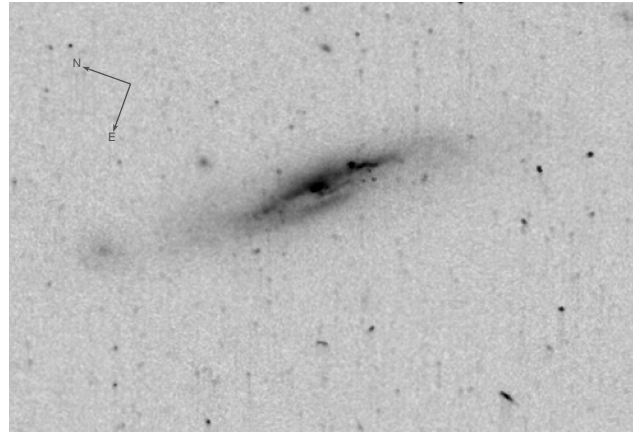}
\caption{A 30$''$x45$''$ HST F450W image of J1703+2451 \protect\citep{Treu:2011aa}. The lens features are not immediately obvious; however, by subtracting the lensing galaxy by a 180$^{\circ}$ rotated version of itself, lensing features appear clearly. This is then used as a mask and fitted to the data to obtain an accurate lens model \protect\citep[see Fig. 4 of][]{Brewer:2012aa}.  }
\label{J1703_HST}
\end{figure}

An HST image of J1703+2451 is shown in Figure \ref{J1703_HST}. This source was originally observed at the VLA in 2013 July; however, due to RFI the data were unusable. It was revisited at the GBT where we obtained a very clean spectrum (Fig. \ref{spectra}(d)); only 8.4 per cent of the data were lost to RFI (see Table \ref{rms_results}). An intermittent, strong RFI signal was present at 1336.7 MHz with a width of 0.2 MHz and a period of 12 s which affected some of our data but was easily identified and removed. Narrowband RFI was also present but was likewise easily removed and only effected single channels.  

A clear signal of width $W_{50}=79\pm13\text{ km s}^{-1}$ and $W_{20}=93\pm12\text{ km s}^{-1}$ is seen at $18852\pm12$ km s$^{-1}$ where $W_{50}$ is the width of the line at 50 per cent flux and $W_{20}$ is the width at 20 per cent flux. J1703+2451 has an inclination of $i=53\pm5$ degrees \citep{Treu:2011aa, Brewer:2012aa, Alam:2015aa} yielding a corrected velocity width of $W_\text{20,cor}=116\pm15$ km s$^{-1}$. We fitted the feature using different velocity resolutions and different baselines.  Because the baselines for this object were good, changing the region where the baseline was fitted did not influence the measurement noticeably (differences of $\sim$1 km s$^{-1}$).  Also, by choosing reasonable different edges of the signal and locations for the horns, the output only changed by $\sim$1 km s$^{-1}$ between fits. 

The location of the line corresponds to redshift $z=0.06289\pm0.00004$. This matches extremely well with the stated SDSS redshift of $z_\text{SDSS}=0.06287\pm0.00001$. The signal-to-noise, S/N, was measured to be $\text{S/N}=6.7$ following the convention of \citet{Saintonge:2007aa},

\begin{equation}
\text{S/N} = \frac{F/W_{50}}{\sigma}\left(\frac{W_{50}/2}{20 \text{ km s}^{-1}}\right)^{1/2}.
\label{SN}
\end{equation}

Here, $F$ is the integrated flux in Jy km s$^{-1}$, $\sigma$ is the rms noise, and 20 km s$^{-1}$ is the velocity resolution of our spectrum. This definition of S/N takes into account the fact that for the same peak flux, a broader spectrum has more signal. 

The measured flux across the line is $\int S \text{d}v = 0.10$ Jy km s$^{-1}$ which yields a mass of $M_\text{H\,\textsc{i}}=(1.77\pm0.06^{+0.35}_{-0.75})\times10^9 \text{ M}_{\odot}$. Statistical errors came from error in the measurements of velocity width, redshift, and flux, with the error in velocity width dominating. Systematic errors came from unaccounted for deficiencies in the reduction process in \textsc{gbtidl} which leads to a mass measurement accurate to about 20 per cent. Furthermore, two background galaxies lie within the field of view in our observations (red dot-dashed lines in Fig. \ref{spectra}(d)) and contribute to the systematic uncertainty on the low end.

\subsubsection{RFI dominated sources}
J0841+3824, J1103+5322, J1117+4704, and J1135+3720 were observed for a total of 44 hours; however, their data were rendered unusable due to the presence of both RFI and systematic noise.

\begin{figure}
\begin{center}
\includegraphics[width=\columnwidth]{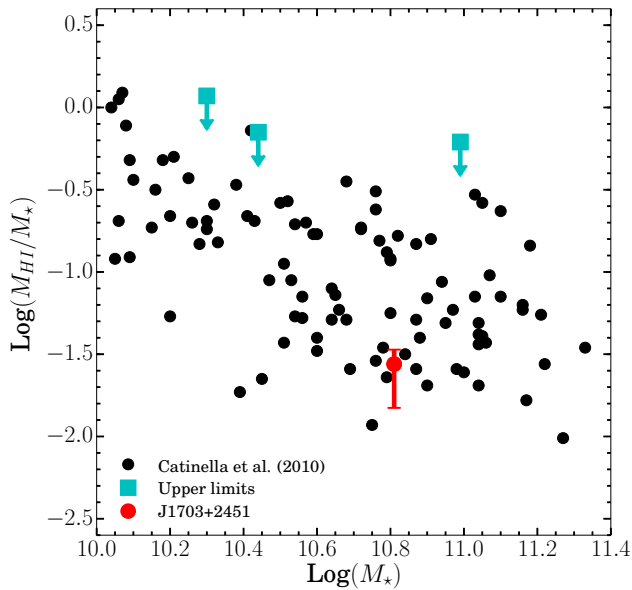}
\caption{Gas fraction versus stellar mass for the galaxies in DR1 of \protect\citet{Catinella:2010aa} and our sources. Upper limits are shown as cyan squares and our detection of J1703+2451 is shown as a red dot. }
\label{GFvsMst}
\end{center}
\end{figure}

\section{Discussion}

\subsection{Background Sources}

As one observes at deeper redshifts, the density of sources increases due to the increased observation volume. Since the field of view for the GBT in the L-band is quite large (9$'$) we must ensure that the only object in our field of view is the target of interest. We therefore checked the SDSS Data Release 12 \citep{Alam:2015aa} database to ensure that the our spectra were not contaminated by background sources. Two of our observations were affected by background sources, J0329-0055 and J1703+2451. 

 Within the field of view of J1703+2451, there are three sources that lie at similar redshift. One is our target and two others are the nearby galaxies SDSS J170334.39+245233.7 (spiral) and SDSS J170332.91+245412.3 (galaxy type uncertain). These two galaxies lie at similar redshifts to our target and are shown as red dot-dashed lines in Fig. \ref{spectra}(d). Good optical spectra exist for both background galaxies which show very strong $H_\alpha$ and [O\,\textsc{ii}] lines, indicating that they are star forming galaxies. Because of their strong emission lines, their redshifts are determined very accurately. Both background galaxies lie more than 26$\sigma$ away from the center of the feature, which indicates that the feature is not a detection of the background galaxies. However, since they do lie near the edge of the feature, we estimate their contribution to the signal.
 
 The spectra for both galaxies have been fit using the Portsmouth Spectral Energy Distribution (SED)-fit pipeline developed by \citet{Maraston:2013aa} which measures the stellar mass using a passive or active star-forming SED template. The measured stellar masses using the star-forming template for the interloper galaxies are $\log(M_\star) = 9.24 \text{ M}_\odot$ and $\log(M_\star)=9.27\text{ M}_\odot$ for SDSS J170334.39+245233.7 and SDSS J170332.91+245412.3 respectively. These measurements indicate that the stellar masses of both galaxies are more than an order of magnitude smaller than the stellar mass of our target.

Both galaxies have been observed with \textit{GALEX}, we therefore can estimate the expected H\,\textsc{i} mass contribution from each galaxy using the linear relationship shown in Fig. \ref{NUV_r_DR1}. The NUV --- \textit{r} color is 3.12 and 2.95 for SDSS J170334.39+245233.7 and SDSS J170332.91+245412.3, respectively. This translates to a signal of $S_\text{peak}\sim0.17$ mJy and $S_\text{peak}\sim0.21$ mJy for SDSS J170334.39+245233.7 and SDSS J170332.91+245412.3, respectively. Individually, both signals are less than the measured rms in the spectrum. If the signals were additive, the resultant would be $S_\text{peak}\sim0.38$ mJy which corresponds to $\text{S/N}=1.4$. Thus, the contribution of both background galaxies is indistinguishable from the noise as there are baseline variations of this magnitude and greater already present in the spectrum where there are no known sources. However, we reflect the estimated contribution of these background galaxies in the systematic error. Since the spectrum represents the total power within the field of view, the presence of the two background galaxies would work to lower the mass measurement and therefore only affects the low end of our systematic error. We add in quadrature the estimated mass of the background galaxies with the other systematic errors from reduction. 

J0329-0055 was observed with the VLA with a beam size of 44$''$$\times$28$''$. Within the field of view lies one other galaxy, SDSS J032957.19-005544.7, which is located at 32340 km s$^{-1}$. No NUV data exists for this object so a mass estimate cannot be made and no signal is seen at that location. However, this serendipitous observation allows us to place an upper limit on this galaxy as well. The measured rms in the channels surrounding the source location is $S_\text{rms}=0.48$ mJy which corresponds to an upper mass limit of $M_\text{H\,\textsc{i}}<3.85\times10^{10}\times(\Delta v/300\text{ km s}^{-1})^{1/2}\text{ M}_\odot$. Within a 4.5$'$ radius (i.e. the field of view of the GBT in the L--band) nine galaxies lie at similar redshift, one of which lies at nearly the same redshift as our source. Therefore, this object would not benefit from observations with a single dish radio telescope as all these sources would be observed simultaneously and could not be disentangled. 

\subsection{Characteristics of J1703+2451}

Comparing the width of the J1703+2451 line to the velocity dispersions measured via gravitational lensing and stellar kinematics (Table \ref{SWELLS-sources}), our detection appears narrow. However, the three measurements consider very different regions of the galaxy. For an undisturbed spiral galaxy, we would expect the circular velocity to be constant out to large radius i.e. a flat rotation curve \citep{Rubin:1978aa}. However, J1703+2451 exhibits a warp in the outer edges of the stellar disk \citep{Brewer:2012aa} that may indicate an interaction with a large perturber in the past. Without high resolution observations, we cannot determine how the velocity dispersion of the gas would be affected in the warped disk. It is also worth noting that the lensing models used in \citet{Brewer:2012aa} to measure the velocity dispersion assumed a singular isothermal ellipsoid model which is known to be overly simplistic for spiral galaxies which have multiple components (disk, bulge, and halo) unlike elliptical galaxies (see, for example, \citet[][]{Barnabe:2012aa}, which have a much better treatment for a spiral galaxy lens model). Furthermore, comparing our measurement of J1703+2451 to galaxies in The H\,\textsc{i} Nearby Galaxy Survey \citep[THINGS;][]{Walter:2008aa} and the GASS sample, there are multiple galaxies of similar stellar mass in both surveys that exhibit H\,\textsc{i} emission lines on the order of 100 km s$^{-1}$ in width.

\begin{figure}
\begin{center}
\includegraphics[width=\columnwidth]{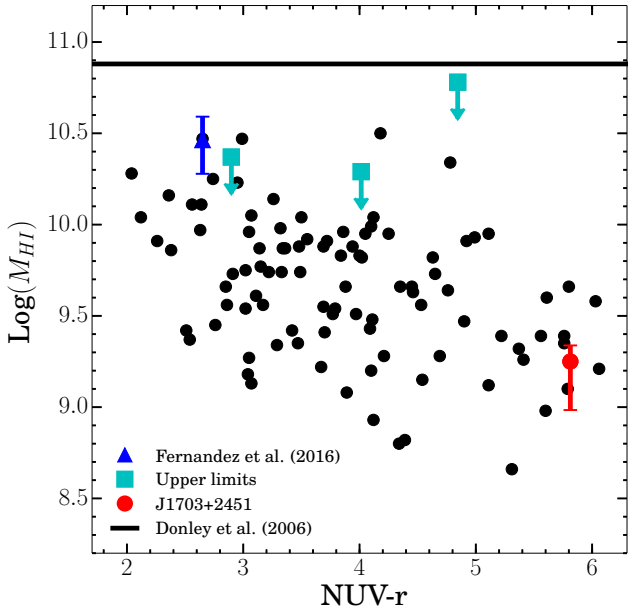}
\caption{Our measured sources versus other published results. The black dots correspond to \protect\citet{Catinella:2010aa} DR1 data, cyan squares correspond to our upper limit measurements, the red dot is our detection of J1703+2451, the blue triangle is the the \protect\citet{Fernandez:2016aa} detection and the solid line is the \protect\citet{Donley:2006aa} detection.}
\label{HIresults}
\end{center}
\end{figure}

Fig. \ref{GFvsMst} shows the relationship between stellar mass and gas fraction for the \citet{Catinella:2010aa} data set and our measurements. Comparing the gas fraction of J1703+2451 to the GASS sample and other galaxies in the Virgo group \citep{Cortese:2011aa}, it appears to lie well within the distribution as an ``H\,\textsc{i} Normal'' galaxy \citep[see Fig. 2 of][]{Huang:2012aa}, meaning that both the measured NUV--r color and gas fraction of J1703+2451 appear to lie along the mean of the GASS distribution of galaxies. 

The uncorrected star formation rate (SFR) of J1703+2451 measured from the strength of the $H_\alpha$ emission line \citep{Kennicutt:1998aa} is SFR $=0.36$ M$_\odot$ yr$^{-1}$. Although this value appears low, the Balmer decrement, which measures the flux ratio of the first two Balmer emission lines and should be equal to 2.85, is measured to be $H_\alpha/H_\beta=6.28$. This indicates a large amount of visual extinction caused by dust. Comparing the uncorrected SFR of J1703+2451 to the uncorrected SFRs of the GASS sample, it appears to lie well within the distribution \citep[see Fig. 4 of][]{Huang:2012aa}. The depletion time for this galaxy is $t_\text{dep}=M_\text{H\,\textsc{i}}/\text{SFR}=4.9$ Gyr, which also aligns well with the GASS sample.

The characteristics of J1703+2451 fit well along the mean of the GASS sample but it is by far our lowest redshift object and lies at the far redshift edge of the GASS sample ($z_\text{GASS}\leq0.05$). The remainder of our sample lies a factor $\sim$2 further in redshift. Due to continuous star formation throughout the lifetime of galaxies, the low redshift sample of galaxies from \citet{Catinella:2010aa} should have a smaller gas reservoir than our target galaxies. On average, $M_\text{H\,\textsc{i}}/M_{\star}$ for the GASS sample is 17 per cent, while gas mass fractions in BzK $z \sim 1$ galaxies are 57 per cent \citep{Daddi:2008aa}, which may be more representative of this sample, as gas fractions are higher at higher redshift \citep{Daddi:2008aa, Tacconi:2010aa}. With this in mind, our mass predictions stated in Table \ref{SWELLS-sources} should be lower than the actual mass of these galaxies. However, both the \citet{Fernandez:2016aa} source and our detection fit along the relation shown in Fig. \ref{NUV_r_DR1}. The relations that are shown are quite wide due to the complicated nature of galaxy evolution and star formation rates, which are affected by galaxy environment, merger timescales, and supernova rates among other things. Therefore, it should be noted that our mass predictions should be used with caution.

In Fig. \ref{HIresults} we see how our results compare to the works of \citet{Donley:2006aa}, \citet{Catinella:2010aa}, and \citet{Fernandez:2016aa}. Due to its location behind the Milky Way, the NUV --- \textit{r} color for the \citet{Donley:2006aa} source is unknown. The SFR for the \citet{Donley:2006aa} source is measured to be 35 $\text{M}_\odot$ yr$^{-1}$, which indicates that it should have strong NUV emission. Although it appears to be very massive compared with the rest of the distribution, it is likely to lie along the general trend and have a low, blueward value for NUV --- \textit{r}. Furthermore, the \citet{Fernandez:2016aa} detection also appears to fit well within the distribution as do a few of our upper limit measurements, notably that of J0329-0055 and our detection of J1703+2451. 

\subsection{Future RFI Management}
Currently, H\,\textsc{i} observations of galaxies at redshifts beyond $z\sim0.1$ are fraught with challenges. For example, the frequency range below 1350 MHz is so filled with RFI signals at Arecibo that it has inhibited our knowledge of the H\,\textsc{i} mass of galaxies beyond redshift $z>0.05$ \citep{Catinella:2010aa, Catinella:2015aa, Giovanelli:2016aa}. Arecibo's superior dish size and sensitivity is negated by the presence of harsh interference signals that are impossible to remove. National Radio Quiet Zones such as the one around GBT help but still do not completely alleviate RFI issues that come from such sources as global positioning satellites and current technology requires many hours of observations for the detection of single galaxies even without interference. Arecibo has been able to detect galaxies at redshifts of $z\sim0.2$; however, due to the large field of view, targets must be selected carefully to lie in low density environments. Next generation H\,\textsc{i} surveys with the SKA and its pathfinders ASKAP (Australian Square Kilometer Array Pathfinder) and MeerKAT (Expanded Karoo Array Telescope) will be able to detect H\,\textsc{i} at much higher redshifts than ever before due to superior spatial resolution and RFI handling \citep{Carilli:2004aa, Johnston:2008aa, Booth:2009aa, Catinella:2015aa, Giovanelli:2016aa}.  Future observatories, such as the SKA, will have better understood systematics and may also employ real-time RFI flagging which will give a more accurate and thorough treatment for RFI affected data \citep{Dumez-Viou:2016aa, van-Nieuwpoort:2017aa}. Furthermore, like the GBT, the SKA will be located in a radio quiet zone and therefore mainly affected only by satellite and airplane interference.  These interferometric surveys will be able to study to H\,\textsc{i} morphology and kinematics of thousands of galaxies across a much wider redshift, shedding light on the H\,\textsc{i} mass function and how the gas supply of spiral galaxies evolves through cosmic time.

\section{Conclusions}

In summary, based on the relation between NUV --- \textit{r} color and the $M_\text{H\,\textsc{i}}/M_\star$ gas fraction, we computed the expected total H\,\textsc{i} masses of the SWELLS sources. We then carried out an observing campaign to secure the H\,\textsc{i} masses of these galaxies. We successfully detected one of our sources, J1703+2451, with a $S/N=6.7$ detection, which is the first detection of H\,\textsc{i} emission in a strong spiral lens. We found a mass of $M_\text{H\,\textsc{i}}= (1.77\pm0.06^{+0.35}_{-0.75})\times10^9 \text{ M}_{\odot}$ with our systematic error reflecting the possible influence of nearby background galaxies that were in the field of view of our observations. The width of the detected H\,\textsc{i} emission line was found to be narrow ($W_{50}=79\pm13$ km s$^{-1}$); however, the SFR, color, gas depletion timescale, and gas fraction of J1703+2451 are all typical for galaxies of its size and redshift. Our upper limits suggest that the other galaxies in this sample that we targeted also have normal gas fractions.

We expect that if RFI and sensitivity are accounted for, H\,\textsc{i} measurements (spectra and maps) should be possible for some of the other SWELLS galaxies. We found significant upper limit H\,\textsc{i} mass constraints of three other galaxies, J0329-0055, J0841+4847, and J1037+3517 and also report a significant upper limit for background galaxy SDSS J032957.19-005544.7 which was serendipitously observed with J0329-0055.  For four other sources, J1111+2234, J1117+4704, J1135+3720, and J1103+5322, our observations were completely unusable due to the presence of strong RFI. 

One of the lessons learned through these observations is that the expected H\,\textsc{i} mass and signal of a galaxy is not necessarily the best factor when prioritizing observations, nor does a large H\,\textsc{i} mass make a target ideal for future follow-up observations. The RFI environment of both single dish and interferometric telescopes must be considered carefully since the RFI environment around each telescope is unique.  Mitigation in the form of nighttime observing can help with FAA radar and cellular RFI signals; however, many RFI issues come from unknown sources and likely from the instrument itself.   Our study indicates that spiral lens galaxies should be detectable with future high resolution H\,\textsc{i} mapping studies.  A significant scientific pay-off of these forthcoming studies is that analysis of the H\,\textsc{i} maps of the SWELLS sample should yield an independent constraint on dark matter substructure that can be compared to the lensing analysis. This combination will provide a more complete picture of substructure in these galaxies which will help to resolve some of the outstanding problems the CDM paradigm has at small scales. Furthermore, it will give us a better understanding of how substructure evolves through cosmic time.

\section*{Acknowledgements}
We are grateful for a careful reading of our paper and many helpful comments from an anonymous referee.
The authors would like to thank Jay Lockman, Andrew Robinson, and Jeyhan Kartaltepe for helpful discussions. AL is supported by the National Radio Astronomy Observatory under the Student Observing Support program project GBT 16B-285. SC acknowledges support from National Science Foundation grant number 1517488. YS is partially supported by the National Natural Science Foundation of China (NSFC) grant 11603032. The authors would like to thank the Sloan Digital Sky Survey for making all their data public and easily accessible and Barbara Catinella and the GASS collaboration for making their data public. The National Radio Astronomy Observatory is a facility of the National Science Foundation operated under cooperative agreement by Associated Universities, Inc. Some of the data presented in this paper were obtained from the Mikulski Archive for Space Telescopes (MAST). STScI is operated by the Association of Universities for Research in Astronomy, Inc., under NASA contract NAS5-26555. Support for MAST for non-HST data is provided by the NASA Office of Space Science via grant NNX09AF08G and by other grants and contracts.




\bibliographystyle{mnras}
\bibliography{thebib}





\bsp	
\label{lastpage}
\end{document}